\documentclass[10pt, twocolumn,showpacs,preprintnumbers,amsmath,amssymb, aip, apl]{revtex4-1}

\usepackage{dcolumn}
\usepackage{bm}

\usepackage[english]{babel}
\usepackage[latin1]{inputenc}
\usepackage[T1]{fontenc}
\usepackage[dvips,final]{graphicx}
\usepackage{epsfig}
\usepackage{amsfonts}
\usepackage[dvips]{color}

\begin{document}

\title{\large{\bf{A micropillar for cavity optomechanics}}}

\affiliation{ONERA, Physics Department; 29 avenue de la Division Leclerc, 92322 Châtillon, France}
\author{A. G. Kuhn}
\email{aurelien.kuhn@spectro.jussieu.fr}
\affiliation{Laboratoire Kastler Brossel, ENS, UPMC, CNRS; case 74, 4 place Jussieu, 75005 Paris, France}
\author{M. Bahriz}
\author{O. Ducloux}
\author{C. Chartier}
\author{O. Le Traon}
\email{Olivier.Le_Traon@onera.fr}
\affiliation{ONERA, Physics Department; 29 avenue de la Division Leclerc, 92322 Châtillon, France}
\author{T. Briant}
\author{P.-F. Cohadon}
\author{A. Heidmann}
\affiliation{Laboratoire Kastler Brossel, ENS, UPMC, CNRS; case 74, 4 place Jussieu, 75005 Paris, France}
\author{C. Michel}
\author{L. Pinard}
\author{R. Flaminio}
\affiliation{Laboratoire des Matériaux Avancés, CNRS, IN2P3; Bâtiment Virgo, 7 avenue Pierre de Coubertin, 69622 Villeurbanne, France}


\begin{abstract}
We present a new micromechanical resonator designed for cavity
optomechanics. We have used a micropillar geometry to obtain a
high-frequency mechanical resonance with a low effective mass and a
very high quality factor. We have coated a 60-$\mu$m diameter
low-loss dielectric mirror on top of the pillar and are planning to
use this micromirror as part of a high-finesse Fabry-Perot cavity,
to laser cool the resonator down to its quantum ground state and to
monitor its quantum position fluctuations by quantum-limited optical
interferometry.
\end{abstract}

\maketitle

Reaching the quantum ground state of a macroscopic mechanical
object is a major experimental challenge in physics, at the origin
of the rapid emergence of the cavity optomechanics research field.
Many groups have been targeting this objective
 for a decade\cite{Roukes,Schwab,GHz,Teufel_QGS}, using a wide range
of resonators oscillating at frequencies from a few Hz to the
GHz-band \cite{Roukes,GHz}, and different techniques of
displacement sensing \cite{Schwab,GHz,tobi,Teufel_strong}. The
development of a very sensitive position sensor combined with a
mechanical resonator in its ground state would have important
consequences\cite{Marquardt}, not only for fundamental aspects in
quantum physics such as entanglement\cite{oa10} and decoherence of
mechanical resonators, but also for potential applications such as
the detection of very weak forces.

Two conditions have to be fulfilled in order to reach and
demonstrate the mechanical ground state. The thermal energy has to
be small with respect to the zero-point quantum energy:
$k_{\rm{B}}T_{\rm c}\ll  h \nu_{\rm m}$. For a resonator
oscillating at a frequency $\nu_{\rm m}=4\,{\rm MHz}$, the
resulting temperature $T_{\rm c}$ is in the sub-mK range and
conventional cryogenic cooling has therefore to be combined with
novel cooling mechanisms such as cavity cooling in a Fabry-Perot
cavity \cite{Teufel_QGS,nat-OA-06, gigan,rsb}.

The second requirement is obviously to be able to detect the very
small residual displacement fluctuations associated with the
quantum ground state. The measurement sensitivity must be better
than the expected displacement noise at resonance, which scales as
$$\textrm{S}_{x}[\nu_{\rm m}] \simeq \left(\frac{25\,µ\textrm{g}}
{M}\right)\left( \frac{Q_{\rm c}}{2\times 10^3}\right) \left(
\frac{4\,\textrm{MHz}}{\nu_{\rm m}}\right)^2
10^{-38}\,\textrm{m}^2/\textrm{Hz},$$
where $M$ is the effective mass of the resonator and $Q_{\rm c}$
its mechanical quality factor. As all optical cooling mechanisms
increase the damping, $Q_{\rm c}$ is here the final quality factor
related to the intrinsic quality factor $Q$ of the resonator by
$Q_{\rm c}T_{\rm c}=QT$, where $T_{\rm c}/T$ is the cooling ratio.
One thus may use a resonator with an ultra-high $Q$, larger than
$10^6$, at a cryogenic temperature of $100\,{\rm mK}$, in order to
ensure a sufficient $Q_{\rm c}\simeq 1000$ at the final effective
temperature.

\begin{figure}
\begin{center}
\includegraphics[width=7cm]{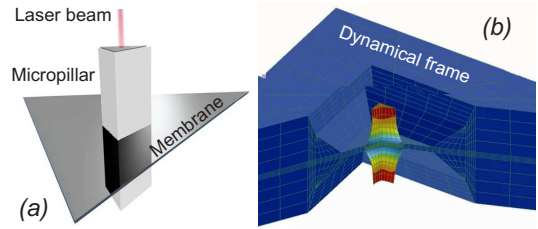}
\caption{3D artist view of the micropillar ({\it a}) and
finite-element simulation of the displacement field of the first
compression-expansion mode of the pillar ({\it b}).} \label{bmp}
\end{center}
\end{figure}

In this letter, we present the mechanical design and experimental
characterization of such a resonator. It consists in a micropillar
mechanically decoupled from the wafer by a dynamical frame and
clamped at its center by a thin membrane. A high-reflectivity
mirror is coated on top of the pillar, thus providing a way for
interferometric sensing of its motion and for future applications
in cavity optomechanics. Advantages of such a pillar geometry are
twofold. First, a compression-expansion mechanical mode of the
pillar has a longitudinal node at its center, decreasing mechanical
loss through the membrane. Second, the top area of the pillar has
a quasi-null strain, reducing the influence of the relatively poor
mechanical quality of the optical layers \cite{LMA} on the overall
mechanical $Q$.

\begin{figure*}[t]
\begin{center}
\includegraphics[width=13cm]{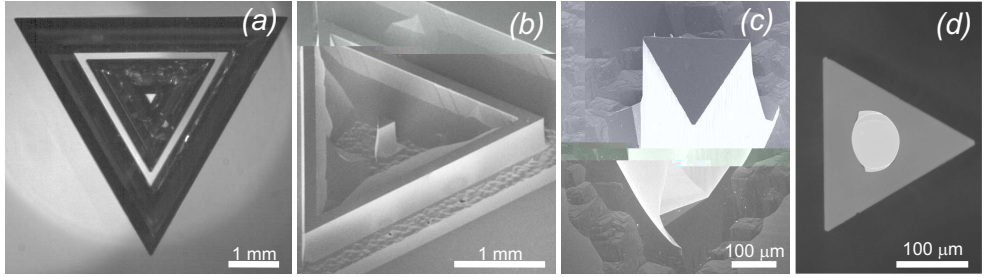}
\caption{Optical view ({\it a}) and scanning electron microscope
picture ({\it b}) of the full structure with the micropillar at
the center, its clamping membrane and dynamical frame; closer view
of the micropillar before the optical coating ({\it c}), and top
view of the coating ({\it d}). \label{Pillar}}
\end{center}
\end{figure*}

We present in the following the full design of the resonator using
finite-element modelling (FEM) simulations, its microfabrication
and experimental mechanical characterization (resonance frequency
and $Q$ factor) using a Michelson-interferometer test bench.

As shown in Fig. \ref{bmp}, the basic structure is a $L=1\,{\rm mm}$
long and 240-$\mu$m wide pillar made of crystalline quartz, clamped
at its center to the external frame by a 20-$\mu$m thin membrane.
According to FEM simulations, with a bulk Young modulus $E =
102.7\,{\rm GPa}$ and a density $\rho = 2648\,{\rm kg/m}^3$,
resonance frequency and effective mass are expected around
$3.2\,{\rm MHz}$ and $25\,\mu{\rm g}$ for the fundamental vibration
mode.

We have chosen to use a single quartz crystal as bulk material to
benefit from its high intrinsic quality factor, with a value
$Q_{\rm int} \simeq 5\times 10^6$ expected at room temperature
\cite{Qint,E}, which can be increased by one to two orders of
magnitude at cryogenic temperature \cite{olt}. Viscous air damping
might be another limiting factor, but is experimentally negligible
for pressures below $10^{-1}\,{\rm mBar}$.

As the effect of the mirror coated on top of the pillar is
expected to be negligible as well, the main limitation to the
quality factor is due to Poisson effects induced inside the
membrane by the radial deformations of the pillar at the vibration
node. A crystallographic orientation of the pillar along the $Z$
axis has been chosen, as well as a pillar equilateral
cross-section with respect to the quartz trigonal symmetry: a
nominally symmetric structure with regard to its median plane is
thus possible, which is a \emph{sine qua non} condition to obtain
a good balancing of the length extension mode. In order to reduce
mechanical losses, we have carefully designed an external
dynamical frame\cite{patent} clamped to the wafer by another
membrane (see Figs. \ref{bmp}\emph{b}, \ref{Pillar}\emph{a} and
\ref{Pillar}\emph{b}). The oscillation of the frame with an
opposite phase then makes up for Poisson effects. The whole
geometry has been optimized by FEM simulation (using SAMCEF code)
to get an exact matching between the pillar and frame momenta,
thus strongly decreasing the energy density at the location of the
external membrane, and reducing clamping losses to a minimum.
Assuming that all the energy remaining within the external
membrane is lost, one gets an underestimated value of $Q$ around
$10^6$.

The resonator microfabrication process takes advantage of standard
techniques of quartz wet etching, using fluorhydric acid. Starting
from a superpolished (down to $2\,$\AA\,rms) ultrapure quartz
substrate, the wafer is metallized with a 15-nm thick layer of
chromium and a 200-nm layer of gold. These layers are used as a
long-time resist mask for the etching of the membrane, down to a
thickness as low as $20\,\mu{\rm m}$. Wet etching ensures quasi
perfect symmetry of the pillar with respect to the membrane, which
appears as an essential condition to reduce clamping losses and to
reach a high $Q$.

Another important condition to reduce mechanical loss is to coat
the mirror only on top of the pillar. For that purpose we use a 3D
5-$\mu$m thick photoresist mask. A photolithography followed by a
developer bath allows one to remove the resist mask over a
well-defined zone, which is a  60-$\mu$m diameter spot centered on
the top of the pillar. The mirror coated by evaporation technique
is a stack of 15 ${\rm SiO}_2/{\rm Ta}_2{\rm O}_5$ quarter-wave
doublets, with a $100\,{\rm ppm}$ expected transmission. After
coating, the photoresist is removed by standard lift-off
techniques, without damaging the optical properties (see Fig.
\ref{Pillar}\emph{d}). Measurement of the micropillar reflectivity
using a dedicated Fabry-Perot cavity is in progress.

\begin{figure}
\begin{center}
\includegraphics[width=7cm]{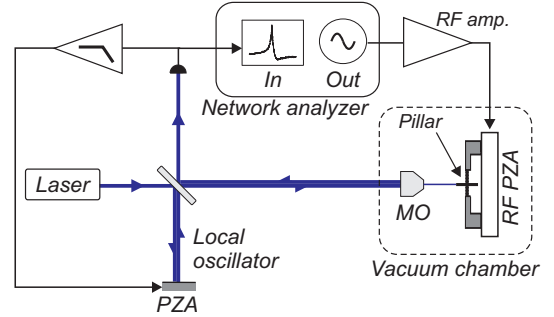}
\caption{Michelson interferometer setup used to characterize the
mechanical properties of the resonator.} \label{mich}
\end{center}
\end{figure}

To characterize the mechanical properties of the resonator, a
Michelson interferometer with a Nd:YAG laser source has been built
(see Fig. \ref{mich}). The micropillar is mounted on a
high-frequency piezoelectric actuator (RF PZA) driven by a
high-power RF amplifier. A microscope objective (MO) focuses the
laser beam down to a few microns on the sample surface. The whole
sensor arm is in a vacuum chamber pumped down to $10^{-3}\,{\rm
mBar}$. A photodiode detects the laser intensity at the output of
the interferometer, its low-frequency signal being used to control
the length of the local oscillator arm through a feedback loop.
The spectral mechanical response of the micropillar is obtained
using a network analyzer to synchronously drive the resonator into
motion and monitor the radio-frequency photodiode signal.

\begin{figure}
\begin{center}
\includegraphics[width=8cm]{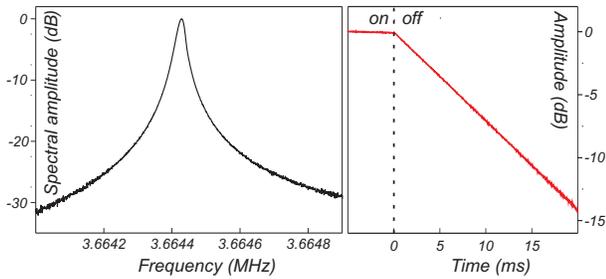}
\caption{Spectrum of the micropillar mechanical response around
its fundamental resonance frequency at $3.66\,{\rm MHz}$ (left),
and ring-down response for the same mechanical mode (right).}
\label{measures}
\end{center}
\end{figure}

Fig. \ref{measures} (left) presents an example of such a
mechanical response spectrum: the mechanical resonance is observed
around $3.66\,{\rm MHz}$, in good agreement with the FEM
simulation results. The slight asymmetry of the resonance is
attributed to an electrical interference between the
optomechanically induced RF signal detected on the photodiode and
a spurious modulation radiated by the RF amplifier. The
interferometric measurement being sensitive to longitudinal
displacements of the surface, only a few and well-isolated
compression-expansion mechanical resonances are observed as
expected from FEM simulations, with a dynamical response at least
$40\,{\rm dB}$ over the mechanical background.

The optical spot can be swept over the whole surface of the
resonator, allowing for a mapping of the vibration profile. It is
mostly found uniform over the upper surface of the pillar, with a
typical displacement amplitude of the order of $5\,{\rm nm}$, to
be compared to the piezoelectric actuator displacement, measured
at a $20\,{\rm pm}$ level on the substrate.

As the intrinsic interferometer jitter may widen the observed
mechanical resonance, its quality factor is measured using a
ring-down technique: once the resonator is driven close to its
resonance frequency, the actuation is switched off and we record
the time evolution of the free resonator motion using a spectrum
analyzer in zero-span mode. The observed exponential decay then
gives the damping time, hence the quality factor. Results depend
on the residual thickness of the membrane: typical values are
larger than $10^5$, reaching values as large as $Q=1.8\times 10^6$
for the best samples without optical coating. Work is in progress
to reduce the membrane thickness and to improve the pillar
symmetry, in particular by using optical coatings on both sides.

To conclude, we have developed a new quartz resonator in a
compression-expansion mode, suitable for cavity optomechanics.
Preliminary mechanical and optical characterizations yield
promising results for the optomechanical coupling that can be
obtained with this device. Next steps will be the operation of the
resonator in a dedicated high-finesse Fabry-Perot cavity, inside a
dilution fridge working at a base temperature of $30\,{\rm mK}$.
Thanks to the very high mechanical quality factor of the
oscillator, we expect to reach a final effective temperature at
the $100\,\mu{\rm K}$ level using laser cooling, thus
demonstrating the quantum ground state of such a macroscopic
optomechanical system.

The authors acknowledge practical help from J. Teissier for the
scanning electron microscope pictures, and financial support of
the "Agence Nationale de la Recherche" (ANR) France, Programme
blanc, N° ANR-07-BLAN-2060 {\it ARQOMM} and of FP7 Specific
Targeted Research Project {\it Minos}.

\end{document}